\documentclass[twocolumn,showpacs,preprintnumbers,amsmath,amssymb]{revtex4}

\usepackage{graphicx}
\usepackage{dcolumn}
\usepackage{bm}

\newcommand{\bq}{\begin{equation}}
\newcommand{\eq}{\end{equation}}
\newcommand{\bqa}{\begin{eqnarray}}
\newcommand{\eqa}{\end{eqnarray}}
\newcommand{\nn}{\nonumber \\}

\def\be     {\begin{equation}}
\def\ee     {\end{equation}}
\def\bea        {\begin{eqnarray}}
\def\eea        {\end{eqnarray}}
\def\bnn    {\begin{eqnarray*}}
\def\enn    {\end{eqnarray*}}

\begin{document}

\title{Role of vertex corrections in the $T$-linear resistivity at the Kondo breakdown quantum critical point}
\author{Ki-Seok Kim}
\affiliation{ Asia Pacific Center for Theoretical Physics, Hogil
Kim Memorial building 5th floor, POSTECH, Hyoja-dong, Namgu,
Pohang, Gyeongbuk 790-784, Korea
\\ Department of Physics, POSTECH, Pohang, Gyeongbuk 790-784, Korea }
\date{\today}

\begin{abstract}
The Kondo breakdown scenario has been claimed to allow the
$T$-linear resistivity in the vicinity of the Kondo breakdown
quantum critical point, two cornerstones of which are the
dynamical exponent $z = 3$ quantum criticality for hybridization
fluctuations in three dimensions and irrelevance of vertex
corrections for transport due to the presence of localized
electrons. We revisit the issue of vertex corrections in
electrical transport coefficients. Assuming that two kinds of
bosonic degrees of freedom, hybridization excitations and gauge
fluctuations, are in equilibrium, we derive coupled quantum
Boltzmann equations for two kinds of fermions, conduction
electrons and spinons. We reveal that vertex corrections play a
certain role, changing the $T$-linear behavior into $T^{5/3}$ in
three dimensions. However, the $T^{5/3}$ regime turns out to be
narrow, and the $T$-linear resistivity is still expected in most
temperature ranges at the Kondo breakdown quantum critical point
in spite of the presence of vertex corrections. We justify our
evaluation, showing that the Hall coefficient is not renormalized
to remain as the Fermi-liquid value at the Kondo breakdown quantum
critical point.
\end{abstract}


\maketitle

\section{Introduction}

It is a long standing problem to understand non-Fermi liquid
transport in condensed matter physics \cite{HF_Review}. In
particular, the mechanism of the $T$-linear resistivity is at the
heart of heavy fermion quantum criticality \cite{T_resistivity},
implying the absence of electron resonances due to strong
inelastic scattering.

A two-dimensional spin-fluctuation scenario demonstrated the
$T$-linear resistivity \cite{Rosch_SDW}. The mechanism is $z = 2$
quantum criticality for spin fluctuations, where $z$ is the
dynamical exponent implying the dispersion of critical
fluctuations. Such two-dimensional fluctuations give rise to the
$T$-linear electron-self-energy. Since vertex corrections are not
relevant due to finite wave-vector ordering, the
temperature-dependence of the relaxation time remains the same as
that of the transport time, resulting in the non-Fermi liquid
resistivity. However, the $T$-linear resistivity results only
within the Eliashberg approximation, where self-energy corrections
for both critical bosons and fermions are introduced but vertex
corrections are neglected \cite{FMQCP}. It was demonstrated that
infinite number of marginal interactions are generated in the two
dimensional $z = 2$ critical theory due to the presence of the
Fermi surface \cite{Abanov}. As a result, logarithmic corrections
due to marginal interactions were argued to give a novel critical
exponent for critical spin dynamics. Then, the self-energy
correction for fermion dynamics may be altered due to the modified
spin dynamics beyond the Eliashberg approximation. It is not clear
at all whether the $T$-linear resistivity is fundamental or not in
the two-dimensional spin-fluctuation scenario. Furthermore, this
mechanism fails to explain the anomalous critical exponent $2/3$
of the Gr\"uneisen ratio in YbRh$_{2}$Si$_{2}$
\cite{GR_Exp,GR_Theory} even within the Eliashberg approximation.

A scenario based on breakdown of the Kondo effect
\cite{KB_Senthil,KB_Indranil,KB_Pepin} has been claimed to cause
the $T$-linear resistivity near the quantum critical point of
YbRh$_{2}$Si$_{2}$ \cite{KB_Indranil,KB_Pepin}. An essential
aspect is that critical hybridization fluctuations are described
by $z = 3$ due to Fermi surface fluctuations of conduction
electrons and localized fermions, giving rise to the $T$-linear
self-energy correction for electron dynamics in three dimensions.
Since the Kondo breakdown transition is involved with zero
momentum ordering, vertex corrections are expected to turn the
$T$-linear relaxation rate into another for the backscattering
rate. However, it was argued that the presence of localized
fermions leads vertex corrections to be irrelevant because
scattering of conduction electrons with hybridization fluctuations
always involves localized fermions and such heavy fermions allow
backscattering to be dominant. As a result, the relaxation rate is
identified with the $T$-linear resistivity of YbRh$_{2}$Si$_{2}$
\cite{Kim_TR}. At the same time, the $z = 3$ quantum criticality
could explain the $2/3$ exponent of the Gr\"uneisen ratio
\cite{Kim_GR}.

In this study we revisit the issue of vertex corrections for the
$T$-linear resistivity of the Kondo breakdown scenario. Assuming
that both hybridization and gauge fluctuations are in equilibrium,
we derive coupled quantum Boltzmann equations for both conduction
electrons and spinons. In contrast with the previous claim on the
irrelevance of vertex corrections for the $T$-linear resistivity
\cite{KB_Indranil,KB_Pepin}, we reveal that vertex corrections
play a certain role, changing the $T$-linear behavior into
$T^{5/3}$ in three dimensions. However, the $T^{5/3}$ regime turns
out to be narrow, and the $T$-linear resistivity is still expected
in most temperature ranges at the Kondo breakdown quantum critical
point in spite of the presence of vertex corrections. We also
calculate the Hall coefficient at the Kondo breakdown quantum
critical point, and find that it is not renormalized because both
longitudinal and transverse resistivities are renormalized by
vertex corrections at the same time.

\section{Kondo breakdown theory}

We start from an effective Anderson lattice model, \bqa && L =
\sum_{i} c_{i\sigma}^{\dagger}(\partial_{\tau} - \mu)c_{i\sigma} -
t \sum_{\langle ij \rangle} (c_{i\sigma}^{\dagger}c_{j\sigma} +
H.c.) \nn && + V \sum_{i} (d_{i\sigma}^{\dagger}c_{i\sigma} +
H.c.) + \sum_{i}d_{i\sigma}^{\dagger}(\partial_{\tau} +
\epsilon_{f})d_{i\sigma} \nn && + J \sum_{\langle ij \rangle}
\vec{S}_{i}\cdot\vec{S}_{j} , \eqa which shows competition between
the Kondo effect ($V$) and the Ruderman-Kittel-Kasuya-Yosida
(RKKY) interaction ($J$). $c_{i\sigma}$ represents an electron in
the conduction band with its chemical potential $\mu$ and hopping
integral $t$. $d_{i\sigma}$ denotes an electron in the localized
orbital with an energy level $\epsilon_{f}$. The localized orbital
experiences strong repulsive interactions, thus either
spin-$\uparrow$ or spin-$\downarrow$ electrons can be occupied at
most. This constraint is incorporated in the U(1) slave-boson
representation, where the localized electron is decomposed into
the holon and spinon, $d_{i\sigma} = b_{i}^{\dagger} f_{i\sigma}$,
supported by the single-occupancy constraint $b_{i}^{\dagger}b_{i}
+ f_{i\sigma}^{\dagger} f_{i\sigma} = S N$ in order to preserve
the physical space. $S = 1/2$ is the size of spin and $N$ is the
spin degeneracy, where the physical case is $N = 2$.

Resorting to the U(1) slave-boson representation, we rewrite the
Anderson lattice model in terms of holons and spinons, \bqa && Z =
\int D c_{i\sigma} D f_{i\sigma} D b_{i} D \chi_{ij} D \lambda_{i}
e^{-\int_{0}^{\beta} d \tau L} , \nn && L = \sum_{i}
c_{i\sigma}^{\dagger}(\partial_{\tau} - \mu)c_{i\sigma} - t
\sum_{\langle ij \rangle} (c_{i\sigma}^{\dagger}c_{j\sigma} +
H.c.) \nn && + \frac{V}{\sqrt{N}} \sum_{i}
(b_{i}f_{i\sigma}^{\dagger}c_{i\sigma} + H.c.) +
\sum_{i}b_{i}^{\dagger} \partial_{\tau} b_{i} \nn && +
\sum_{i}f_{i\sigma}^{\dagger}(\partial_{\tau} +
\epsilon_{f})f_{i\sigma} - J \sum_{\langle ij \rangle} (
f_{i\sigma}^{\dagger}\chi_{ij}f_{j\sigma} + H.c.) \nn && + i
\sum_{i} \lambda_{i} (b_{i}^{\dagger}b_{i} + f_{i\sigma}^{\dagger}
f_{i\sigma} - S N) + N J \sum_{\langle ij \rangle} |\chi_{ij}|^{2}
, \eqa where the RKKY spin-exchange term for the localized orbital
is decomposed via exchange hopping processes of spinons with a
hopping parameter $\chi_{ij}$, and $\lambda_{i}$ is a Lagrange
multiplier field to impose the single-occupancy constraint.

The saddle-point analysis with $b_{i} \rightarrow b$, $\chi_{ij}
\rightarrow \chi$, and $i\lambda_{i} \rightarrow \lambda$ reveals
breakdown of the Kondo effect \cite{KB_Senthil}, where a
spin-liquid Mott insulator ($b = 0$) arises with a small area of
the Fermi surface in $J > T_{K}$ while a heavy Fermi liquid ($b
\not= 0$) obtains with a large Fermi surface in $T_{K}
> J$. Here, $T_{K} = D \exp\Bigl(\frac{\epsilon_{f}}{N
\rho_{c}V^{2}}\Bigr)$ is the single-ion Kondo temperature, where
$\rho_{c} \approx (2D)^{-1}$ is the density of states for
conduction electrons with the half bandwidth $D$. Reconstruction
of the Fermi surface occurs at $J \simeq T_{K}$.

Quantum critical physics is characterized by critical fluctuations
of the hybridization order parameter, introduced in the Eliashberg
theory \cite{FMQCP}, where self-energy corrections of electrons,
spinons, and holons are taken into account fully self-consistently
but vertex corrections are not incorporated \cite{Kim_LW}.
Dynamics of critical Kondo fluctuations is described by $z = 3$
critical theory due to Landau damping of electron-spinon
polarization above an intrinsic energy scale $E^{*}$, while by $z
= 2$ dilute Bose gas model below $E^{*}$
\cite{KB_Indranil,KB_Pepin}. The energy scale $E^{*}$ originates
from the mismatch of Fermi surfaces of conduction electrons and
spinons, one of the central aspects in the Kondo breakdown
scenario. Physically, one may understand that quantum fluctuations
of the Fermi-surface reconfiguration start to be frozen at $T
\approx E^{*}$, thus the conduction electron's Fermi surface
dynamically decouples from the spinon's one below $E^{*}$. The
Kondo breakdown scenario claimed that such an energy scale was
actually measured in the Seebeck coefficient, interpreting an
abrupt collapse of the Seebeck coefficient to result from the
decoupling effect of Fermi surfaces \cite{Kim_SeeBeck}.

\section{Quantum Boltzmann equation approach}

\begin{widetext}
Based on the effective field theory [Eq. (2)], we evaluate both
longitudinal and transverse transport coefficients. We start from
coupled quantum Boltzman equations, given by \bqa && \frac{e}{c}
\boldsymbol{v}_{k}^{c(f)} \cdot (\boldsymbol{B}_{c(f)} \times
\boldsymbol{\partial}_{\boldsymbol{k}}) G_{c(f)}^{<}(k,\omega) + e
\boldsymbol{E}_{c(f)} \cdot \boldsymbol{v}_{k}^{c(f)}
[\partial_{\omega} f(\omega)] \Gamma_{c(f)}(k,\omega)
[A_{c(f)}(k,\omega)]^{2} = I_{coll}^{c(f)}(k,\omega) , \nn &&
I_{coll}^{c(f)}(k,\omega) = 2i \Gamma_{c(f)}(k,\omega)
G^{<}_{c(f)}(k,\omega) - i \Sigma^{<}_{c(f)}(k,\omega)
A_{c(f)}(k,\omega) \eqa for conduction electrons (spinons), where
$G_{c(f)}^{<}(k,\omega)$ and $\Sigma_{c(f)}^{<}(k,\omega)$ are
lesser Green's function and self-energy of conduction electrons
(spinons), respectively, and $A_{c(f)}(k,\omega)$ and
$\Gamma_{c(f)}(k,\omega)$ are imaginary parts of retarded Green's
function and self-energy, respectively.
$\boldsymbol{v}_{k}^{c(f)}$ is the velocity of electrons
(spinons). $f(\omega)$ is the Fermi-Dirac distribution function.
$\boldsymbol{E}_{c} = \boldsymbol{E}$ and $\boldsymbol{B}_{c} =
\boldsymbol{B}$ are applied electric and magnetic fields while
$\boldsymbol{E}_{f} = \boldsymbol{\mathcal{E}}$ and
$\boldsymbol{B}_{f} = \boldsymbol{\mathcal{B}}$ are internal
fields related with fractionalization. Since spinons do not carry
an electric charge in our assignment, they couple to internal
fields only in a gauge invariant way. Derivation of these
equations is presented in Ref. \cite{QBE_MIT}.

Inelastic scattering with critical fluctuations gives rise to the
collision term of the right-hand-side, where each lesser
self-energy is given by $\Sigma_{c}^{<}(k,\omega) =
\Sigma_{c}^{b<}(k,\omega)$ and $\Sigma_{f}^{<}(k,\omega) =
\Sigma_{f}^{b<}(k,\omega) + \Sigma_{f}^{a<}(k,\omega)$, \bqa &&
\Sigma^{b<}_{c(f)}(k,\omega) = V^{2} \sum_{q} \int_{0}^{\infty}
\frac{d\nu}{\pi} \Im D_{b}(q,\nu) [\{ n(\nu) + 1\}
G^{<}_{f(c)}(k+q,\omega+\nu) + n(\nu)
G^{<}_{f(c)}(k+q,\omega-\nu)] , \nn && \Sigma_{f}^{a<}(k,\omega) =
\sum_{q} \int_{0}^{\infty} \frac{d\nu}{\pi} \Bigl|
\frac{k\times\hat{q}}{m_{f}} \Bigr|^{2} \Im D_{a}(q,\nu) [\{
n(\nu) + 1\} G^{<}_{f}(k+q,\omega+\nu) + n(\nu)
G^{<}_{f}(k+q,\omega-\nu)] .   \eqa The superscript, $b$ or $a$,
means the scattering source, corresponding to either hybridization
fluctuations or gauge excitations. Although scattering of spinons
with gauge fluctuations was not emphasized in the previous
section, such fluctuations represent certain types of collective
spin fluctuations associated with spin chirality
\cite{Lee_Nagaosa}, and they contribute to non-Fermi liquid
physics. See Ref. \cite{Kim_GR} in order to understand how much
they contribute to thermodynamics at the Kondo breakdown quantum
critical point. $\Im D_{b(a)}(q,\nu) \propto \gamma_{b(a)} q \nu /
(q^{6} + \gamma_{b(a)}^{2} \nu^{2})$ is the spectral function of
the hybridization-fluctuation (gauge) propagator, given by $z = 3$
in the quantum critical regime, where $\gamma_{b(a)}$ is the
Landau damping constant \cite{KB_Indranil,KB_Pepin}.


Inserting the lesser Green's functions \bqa G^{<}_{c(f)}(k,\omega)
&=& i A_{c(f)}(k,\omega) \Bigl\{ f(\omega) + \Bigl( -
\frac{\partial f(\omega)}{\partial \omega} \Bigr)
\boldsymbol{v}_{k}^{c(f)} \cdot
\boldsymbol{\Lambda}_{c(f)}(k,\omega) \Bigr\}
%
%
%
\eqa into the quantum Boltzman equations for both conduction
electrons and spinons, we obtain \bqa && i \frac{e}{c} [
\boldsymbol{v}_{\boldsymbol{k}}^{c} \cdot (\boldsymbol{B} \times
\boldsymbol{\partial}_{\boldsymbol{k}})
\boldsymbol{v}_{\boldsymbol{k}}^{c} ] \cdot
\boldsymbol{\Lambda}_{c}(k,\omega) -  e \boldsymbol{E} \cdot
\boldsymbol{v}_{k}^{c} \Gamma_{c}^{b}(k,\omega) A_{c}(k,\omega) =
- 2 \Gamma_{c}^{b}(k,\omega) \boldsymbol{v}_{k}^{c} \cdot
\boldsymbol{\Lambda}_{c}(k,\omega) \nn && + V^{2} \sum_{q}
\int_{0}^{\infty} \frac{d\nu}{\pi} \Im D_{b}(q,\nu) \Bigl\{ \{
n(\nu) + f(\omega+\nu) \} A_{f}(k+q,\omega+\nu)
\boldsymbol{v}_{k+q}^{f} \cdot
\boldsymbol{\Lambda}_{f}(k+q,\omega+\nu) \nn && - \{ n(-\nu) +
f(\omega-\nu) \} A_{f}(k+q,\omega-\nu) \boldsymbol{v}_{k+q}^{f}
\cdot \boldsymbol{\Lambda}_{f}(k+q,\omega-\nu) \Bigr\}  \eqa for
conduction electrons, and \bqa && i \frac{e}{c} [
\boldsymbol{v}_{\boldsymbol{k}}^{f} \cdot
(\boldsymbol{\mathcal{B}} \times
\boldsymbol{\partial}_{\boldsymbol{k}})
\boldsymbol{v}_{\boldsymbol{k}}^{f} ] \cdot
\boldsymbol{\Lambda}_{f}(k,\omega) - e \boldsymbol{\mathcal{E}}
\cdot \boldsymbol{v}_{k}^{f} \Gamma_{f}(k,\omega) A_{f}(k,\omega)
= - 2 \Gamma_{f}(k,\omega) \boldsymbol{v}_{k}^{f} \cdot
\boldsymbol{\Lambda}_{f}(k,\omega) \nn && + V^{2} \sum_{q}
\int_{0}^{\infty} \frac{d\nu}{\pi} \Im D_{b}(q,\nu) \Bigl\{ \{
n(\nu) + f(\omega+\nu) \} A_{c}(k+q,\omega+\nu)
\boldsymbol{v}_{k+q}^{c} \cdot
\boldsymbol{\Lambda}_{c}(k+q,\omega+\nu) \nn && - \{ n(-\nu) +
f(\omega-\nu) \} A_{c}(k+q,\omega-\nu) \boldsymbol{v}_{k+q}^{c}
\cdot \boldsymbol{\Lambda}_{c}(k+q,\omega-\nu) \Bigr\} \nn && +
\sum_{q} \int_{0}^{\infty} \frac{d\nu}{\pi} \Bigl|
\frac{k\times\hat{q}}{m_{f}} \Bigr|^{2} \Im D_{a}(q,\nu) \Bigl\{
\{ n(\nu) + f(\omega+\nu) \} A_{f}(k+q,\omega+\nu)
\boldsymbol{v}_{k+q}^{f} \cdot
\boldsymbol{\Lambda}_{f}(k+q,\omega+\nu) \nn && - \{ n(-\nu) +
f(\omega-\nu) \} A_{f}(k+q,\omega-\nu) \boldsymbol{v}_{k+q}^{f}
\cdot \boldsymbol{\Lambda}_{f}(k+q,\omega-\nu) \Bigr\} \eqa for
spinons. $\boldsymbol{\Lambda}_{c(f)}(k,\omega)$ are
non-equilibrium distribution functions, containing the information
of vertex corrections. Since hybridization fluctuations are
involved with both conduction electrons and spinons, quantum
Boltzmann equations for both distribution functions are coupled.
We show that this coupled dynamics gives rise to nontrivial vertex
corrections in transport coefficients.


In order to solve these coupled equations with magnetic fields, we
rewrite Eqs. (6) and (7) in terms of $x$ and $y$ directions, given
by \bqa && - i \omega_{c} \Lambda_{c}^{y}(k_{F},\omega) - e E_{x}
\Gamma_{c}(k_{F},\omega) A_{c}(k_{F},\omega) = - 2
\Gamma_{c}(k_{F},\omega) \Lambda_{c}^{x}(k_{F},\omega) + V^{2}
\frac{N_{F}^{f}}{2\pi} \int{d\xi} \int_{-1}^{1} d\cos\theta_{cf}
\int_{0}^{\infty} \frac{d\nu}{\pi} \Im D_{b}(q,\nu) \nn && \Bigl\{
\{ n(\nu) + f(\omega+\nu) \} A_{f}(k_{F}+q,\omega+\nu) - \{
n(-\nu) + f(\omega-\nu) \} A_{f}(k_{F}+q,\omega-\nu) \Bigr\}
\Bigl( \frac{v_{F}^{f}} {v_{F}^{c}} \cos\theta_{cf} \Bigr)
\Lambda_{f}^{x}(k_{F},\omega) , \nn && i \omega_{c}
\Lambda_{c}^{x}(k_{F},\omega) - e E_{y} \Gamma_{c}(k_{F},\omega)
A_{c}(k_{F},\omega) = - 2 \Gamma_{c}(k_{F},\omega)
\Lambda_{c}^{y}(k_{F},\omega) + V^{2} \frac{N_{F}^{f}}{2\pi}
\int{d\xi} \int_{-1}^{1} d\cos\theta_{cf} \int_{0}^{\infty}
\frac{d\nu}{\pi} \Im D_{b}(q,\nu) \nn && \Bigl\{ \{ n(\nu) +
f(\omega+\nu) \} A_{f}(k_{F}+q,\omega+\nu) - \{ n(-\nu) +
f(\omega-\nu) \} A_{f}(k_{F}+q,\omega-\nu) \Bigr\} \Bigl(
\frac{v_{F}^{f}} {v_{F}^{c}} \cos\theta_{cf} \Bigr)
\Lambda_{f}^{y}(k_{F},\omega) \eqa for conduction electrons with
the cyclotron frequency $\omega_{c} = \frac{e B}{m_{c} c}$, and
\bqa && - i \Omega_{f} \Lambda_{f}^{y}(k_{F},\omega) - e
\mathcal{E}_{x} \Gamma_{f}(k_{F},\omega) A_{f}(k_{F},\omega) = - 2
\Gamma_{f}(k_{F},\omega) \Lambda_{f}^{x}(k_{F},\omega) + V^{2}
\frac{N_{F}^{c}}{2\pi} \int d\xi \int_{-1}^{1} d\cos\theta_{cf}
\int_{0}^{\infty} \frac{d\nu}{\pi} \Im D_{b}(q,\nu) \nn && \Bigl\{
\{ n(\nu) + f(\omega+\nu) \} A_{c}(k_{F}+q,\omega+\nu) - \{
n(-\nu) + f(\omega-\nu) \} A_{c}(k_{F}+q,\omega-\nu) \Bigr\}
\Bigl( \frac{v_{F}^{c}}{v_{F}^{f}} \cos \theta_{cf} \Bigr)
\Lambda_{c}^{x}(k_{F},\omega) \nn && + \frac{N_{F}^{f}}{2\pi} \int
d\xi \int_{-1}^{1} d\cos\theta_{ff} \int_{0}^{\infty}
\frac{d\nu}{\pi} [v_{F}^{f2} \cos^{2}(\theta_{ff}/2)] \Im
D_{a}(q,\nu) \Bigl\{ \{ n(\nu) + f(\omega+\nu) \}
A_{f}(k_{F}+q,\omega+\nu) \nn && - \{ n(-\nu) + f(\omega-\nu) \}
A_{f}(k_{F}+q,\omega-\nu) \Bigr\} \cos \theta_{ff}
\Lambda_{f}^{x}(k_{F},\omega) , \nn && i \frac{e}{m_{f} c}
\mathcal{B} \Lambda_{f}^{x}(k_{F},\omega) - e \mathcal{E}_{y}
\Gamma_{f}(k_{F},\omega) A_{f}(k_{F},\omega) = - 2
\Gamma_{f}(k_{F},\omega) \Lambda_{f}^{y}(k_{F},\omega) + V^{2}
\frac{N_{F}^{c}}{2\pi} \int d\xi \int_{-1}^{1} d\cos\theta_{cf}
\int_{0}^{\infty} \frac{d\nu}{\pi} \Im D_{b}(q,\nu) \nn && \Bigl\{
\{ n(\nu) + f(\omega+\nu) \} A_{c}(k_{F}+q,\omega+\nu) - \{
n(-\nu) + f(\omega-\nu) \} A_{c}(k_{F}+q,\omega-\nu) \Bigr\}
\Bigl( \frac{v_{F}^{c}}{v_{F}^{f}} \cos \theta_{cf} \Bigr)
\Lambda_{c}^{y}(k_{F},\omega) \nn && + \frac{N_{F}^{f}}{2\pi} \int
d\xi \int_{-1}^{1} d\cos\theta_{ff} \int_{0}^{\infty}
\frac{d\nu}{\pi} [v_{F}^{f2} \cos^{2}(\theta_{ff}/2)] \Im
D_{a}(q,\nu) \Bigl\{ \{ n(\nu) + f(\omega+\nu) \}
A_{f}(k_{F}+q,\omega+\nu) \nn && - \{ n(-\nu) + f(\omega-\nu) \}
A_{f}(k_{F}+q,\omega-\nu) \Bigr\} \cos \theta_{ff}
\Lambda_{f}^{y}(k_{F},\omega) \eqa for spinons with an internal
cyclotron frequency $\Omega_{f} = \frac{e \mathcal{B}}{m_{f} c}$.
In this derivation we perform the following approximation \bqa &&
\boldsymbol{\Lambda}_{c(f)}(k+q,\omega \pm \nu) \approx
\boldsymbol{\Lambda}_{c(f)}(k_{F},\omega) , \eqa regarded as the
zeroth-order. We checked the validity of this approximation,
applying Eq. (10) into two problems such as transport with
impurity scattering and that in the spin liquid state and
recovering known results \cite{QBE_MIT,QBE_KB}. We also recover
the conventional expression in this problem, if vertex corrections
are neglected.

It is straightforward to solve these coupled linear algebraic
equations. Introducing \bqa \Lambda_{c(f)}^{x}(k_{F},\omega) + i
\Lambda_{c(f)}^{y}(k_{F},\omega) = \Lambda_{c(f)}(k_{F},\omega) ,
~~~~~ E_{x} + i E_{y} = E \eqa with the complex notation, we find
non-equilibrium distribution functions \bqa && \Lambda_{c}(k_{F}
,\omega) = e E \frac{\Gamma_{c}^{b}(k_{F} ,\omega) A_{c}(k_{F}
,\omega)}{ 2 \Gamma_{c}^{b}(k_{F} ,\omega) + i \omega_{c} } +
\frac{v_{F}^{f}} {v_{F}^{c}} \frac{2 \Gamma_{c, cos}^{b}(k_{F}
,\omega) }{ 2 \Gamma_{c}^{b}(k_{F} ,\omega) + i \omega_{c}}
\Lambda_{f} (k_{F} ,\omega) \eqa for conduction electrons, and
\bqa && \Lambda_{f} (k_{F} ,\omega) = e \mathcal{E} \frac{[
\Gamma_{f}^{b}(k_{F} ,\omega) + \Gamma_{f}^{a}(k_{F} ,\omega)]
A_{f}(k_{F} ,\omega)}{2 \Gamma_{f}^{b}(k_{F} ,\omega) + 2
\Gamma_{f, tr}^{a}(k_{F} ,\omega) + i \Omega_{f}} +
\frac{v_{F}^{c}}{v_{F}^{f}} \frac{2 \Gamma_{f,cos}^{b}(k_{F}
,\omega)}{2 \Gamma_{f}^{b}(k_{F} ,\omega) + 2 \Gamma_{f,
tr}^{a}(k_{F} ,\omega) + i \Omega_{f}} \Lambda_{c} (k_{F},\omega)
\eqa for spinons, respectively. Scattering with hybridization
fluctuations gives rise to two kinds of scattering rates, \bqa &&
2 \Gamma_{c(f)}^{b}(k_{F} ,\omega) = V^{2}
\frac{N_{F}^{f(c)}}{2\pi} \int{d\xi} \int_{-1}^{1}
d\cos\theta_{cf} \int_{0}^{\infty} \frac{d\nu}{\pi} \Im
D_{b}(q,\nu) \nn && \Bigl\{ \{ n(\nu) + f(\omega+\nu) \}
A_{f(c)}(k_{F}+q,\omega+\nu) - \{ n(-\nu) + f(\omega-\nu) \}
A_{f(c)}(k_{F}+q,\omega-\nu) \Bigr\} , \nn && 2 \Gamma_{c (f),
cos}^{b}(k_{F} ,\omega) = V^{2} \frac{N_{F}^{f(c)}}{2\pi}
\int{d\xi} \int_{-1}^{1} d\cos\theta_{cf} \int_{0}^{\infty}
\frac{d\nu}{\pi} \Im D_{b}(q,\nu) \nn && \Bigl\{ \{ n(\nu) +
f(\omega+\nu) \} A_{f(c)}(k_{F}+q,\omega+\nu) - \{ n(-\nu) +
f(\omega-\nu) \} A_{f(c)}(k_{F}+q,\omega-\nu) \Bigr\}
\cos\theta_{cf} , \eqa where the former corresponds to the
relaxation rate and the latter is associated with the transport
time, denoted from the $\cos \theta_{cf}$ term. $\theta_{cf}$
represents an angle between the Fermi velocity of conduction
electrons and that of spinons. Gauge fluctuations result in
relaxation to spinons, \bqa && 2 \Gamma_{f}^{a}(k,\omega) =
\frac{N_{F}^{f}}{2\pi} \int d\xi \int_{-1}^{1} d\cos\theta_{ff}
\int_{0}^{\infty} \frac{d\nu}{\pi} [v_{F}^{f2}
\cos^{2}(\theta_{ff}/2)] \Im D_{a}(q,\nu) \nn && \Bigl\{ \{ n(\nu)
+ f(\omega+\nu) \} A_{f}(k+q,\omega+\nu) - \{ n(-\nu) +
f(\omega-\nu) \} A_{f}(k+q,\omega-\nu) \Bigr\} , \nn && 2
\Gamma_{f, tr}^{a}(k,\omega) = \frac{N_{F}^{f}}{2\pi} \int d\xi
\int_{-1}^{1} d\cos\theta_{ff} \int_{0}^{\infty} \frac{d\nu}{\pi}
[v_{F}^{f2} \cos^{2}(\theta_{ff}/2)] \Im D_{a}(q,\nu) \nn &&
\Bigl\{ \{ n(\nu) + f(\omega+\nu) \} A_{f}(k_{F}+q,\omega+\nu) -
\{ n(-\nu) + f(\omega-\nu) \} A_{f}(k_{F}+q,\omega-\nu) \Bigr\} (1
- \cos \theta_{ff}) , \eqa where the former is the relaxation rate
and the latter is the backscattering rate, associated with the $1
- \cos \theta_{ff}$ term. $\theta_{ff}$ is an angle between the
Fermi velocities of spinons before and after scattering. We point
out that gauge invariance gives rise to $\Gamma_{f, tr}^{a}(k_{F}
,\omega)$ instead of $\Gamma_{f}^{a}(k_{F} ,\omega)$ in the
denominator of the spinon distribution function \cite{QBE_KB}.
This issue was intensively discussed based on the diagrammatic
approach \cite{YB_Diagram} and the quantum Boltzmann equation
approach \cite{YB_QBE}.

Second terms in Eqs. (12) and (13) result from vertex corrections.
In other words, if such contributions are ignored, we recover the
$T$-linear resistivity as claimed before. Inserting Eq. (12) into
Eq. (13), we obtain the following expression \bqa && \Lambda_{f}
(k_{F} ,\omega) = \frac{ e \mathcal{E} [ \Gamma_{f}^{b}(k_{F}
,\omega) + \Gamma_{f}^{a}(k_{F} ,\omega)] A_{f}(k_{F} ,\omega) + e
E \frac{v_{F}^{c}}{v_{F}^{f}} \frac{ 2 \Gamma_{f,cos}^{b}(k_{F}
,\omega) \Gamma_{c}^{b}(k_{F} ,\omega) }{ 2 \Gamma_{c}^{b}(k_{F}
,\omega) + i \omega_{c} } A_{c}(k_{F} ,\omega) }{2
\Gamma_{f}^{b}(k_{F} ,\omega) + 2 \Gamma_{f, tr}^{a}(k_{F}
,\omega) + i \Omega_{f} - \frac{4 \Gamma_{f,cos}^{b}(k_{F}
,\omega) \Gamma_{c, cos}^{b}(k_{F} ,\omega) }{ 2
\Gamma_{c}^{b}(k_{F} ,\omega) + i \omega_{c}}} \eqa for the spinon
distribution function. Inserting this equation into Eq. (12), we
obtain the distribution function for conduction electrons, \bqa &&
\Lambda_{c}(k_{F} ,\omega) = e E \frac{\Gamma_{c}^{b}(k_{F}
,\omega) A_{c}(k_{F} ,\omega)}{ 2 \Gamma_{c}^{b}(k_{F} ,\omega) +
i \omega_{c} } \nn && + \frac{2 \Gamma_{c, cos}^{b}(k_{F} ,\omega)
}{ 2 \Gamma_{c}^{b}(k_{F} ,\omega) + i \omega_{c}} \frac{ e
\mathcal{E} \frac{v_{F}^{f}} {v_{F}^{c}} [ \Gamma_{f}^{b}(k_{F}
,\omega) + \Gamma_{f}^{a}(k_{F} ,\omega)] A_{f}(k_{F} ,\omega) + e
E \frac{ 2 \Gamma_{f,cos}^{b}(k_{F} ,\omega) \Gamma_{c}^{b}(k_{F}
,\omega) }{ 2 \Gamma_{c}^{b}(k_{F} ,\omega) + i \omega_{c} }
A_{c}(k_{F} ,\omega) }{2 \Gamma_{f}^{b}(k_{F} ,\omega) + 2
\Gamma_{f, tr}^{a}(k_{F} ,\omega) + i \Omega_{f} - \frac{4
\Gamma_{f,cos}^{b}(k_{F} ,\omega) \Gamma_{c, cos}^{b}(k_{F}
,\omega) }{ 2 \Gamma_{c}^{b}(k_{F} ,\omega) + i \omega_{c}}} .
\eqa Recalling $v_{F}^{f} / v_{F}^{c} = \alpha \ll 1$ at the
quantum critical point \cite{Kim_TR}, we approximate the above
expression as follows \bqa &&  \Lambda_{c}(k_{F} ,\omega) \approx
e E \frac{\Gamma_{c}^{b}(k_{F} ,\omega)}{ 2 \Gamma_{c}^{b}(k_{F}
,\omega) + i \omega_{c} } \frac{2 \Gamma_{f}^{b}(k_{F} ,\omega) +
2 \Gamma_{f, tr}^{a}(k_{F} ,\omega) + i \Omega_{f}}{2
\Gamma_{f}^{b}(k_{F} ,\omega) + 2 \Gamma_{f, tr}^{a}(k_{F}
,\omega) + i \Omega_{f} - \frac{4 \Gamma_{f,cos}^{b}(k_{F}
,\omega) \Gamma_{c, cos}^{b}(k_{F} ,\omega) }{ 2
\Gamma_{c}^{b}(k_{F} ,\omega) + i \omega_{c}}} A_{c}(k_{F}
,\omega) . \eqa

Inserting the distribution function into the definition of an
electric current \bqa && J_{c} = J_{c}^{x} + i J_{c}^{y} = e^{2}
v_{F}^{c 2} \int\frac{d^{3}k}{(2\pi)^{3}} \int\frac{d\omega}{2\pi}
\sin^{2} \theta_{k}^{cc} \Bigl(- \frac{\partial
f(\omega)}{\partial \omega} \Bigr) A_{c}(k_{F},\omega)
\Lambda_{c}(k_{F},\omega) ,
%
%
\eqa we obtain \bqa && J_{c} \approx \mathcal{C}_{c} e^{2}
v_{F}^{c 2} \rho_{c} \frac{[\tau_{sc}^{b}(T)]^{-1} +
[\tau_{tr}^{a}(T)]^{-1}}{ [\tau_{sc}(T)]^{-1} \bigl(
[\tau_{tr}^{b}(T)]^{-1} + [\tau_{tr}^{a}(T)]^{-1} \bigr) + i
\omega_{c} \bigl( [\tau_{sc}^{b}(T)]^{-1} +
[\tau_{tr}^{a}(T)]^{-1} \bigr) } E , \eqa where $2
\Gamma_{c}^{b}(T) \rightarrow [\tau_{sc}(T)]^{-1}$, $2
\Gamma_{f}^{b}(T) \rightarrow [\tau_{sc}^{b}(T)]^{-1}$, $2
\Gamma_{f,tr}^{b}(T) \rightarrow [\tau_{tr}^{b}(T)]^{-1}$, and $2
\Gamma_{f,tr}^{a}(T) \rightarrow [\tau_{tr}^{a}(T)]^{-1}$ are
used. We also adopt $\Omega_{f} \ll \omega_{c}$ due to $m_{f} \gg
m_{c}$, where $m_{c(f)}$ is the band mass of conduction electrons
(spinons). $\mathcal{C}_{c}$ is a positive numerical constant,
given by \bqa && \mathcal{C}_{c} = \mathcal{C} \int_{-1}^{1} d
\cos \theta_{k_{F}}^{cc} \sin^{2} \theta_{k_{F}}^{cc}
\int_{-\infty}^{\infty} d \xi \Gamma_{c}^{b} (k_{F},\omega)
[A_{c}(\xi,\omega)]^{2} , \nonumber \eqa where \bqa &&
A_{c}(\xi,\omega) = - \frac{1}{\pi} \frac{\Im
\Sigma_{c}(k_{F},\omega)}{[\omega - \xi - \Re
\Sigma_{c}(k_{F},\omega)]^{2} + [\Im
\Sigma_{c}(k_{F},\omega)]^{2}} \nonumber \eqa is the spectral
function of conduction electrons and $\mathcal{C}$ is a positive
numerical constant appearing from the angle integration.

An approximation is $\Gamma_{c, cos}^{b}(k_{F} ,\omega) \approx
\Gamma_{c}^{b}(k_{F} ,\omega)$ from Eq. (18). An accurate
treatment will be $\Gamma_{c, cos}^{b}(k_{F} ,\omega) =
\Gamma_{c}^{b}(k_{F} ,\omega) - \Gamma_{c, tr}^{b}(k_{F}
,\omega)$. It turns out that this replacement does not modify our
conclusion at all because $\Gamma_{c, tr}^{b}(k_{F} ,\omega)$ is
irrelevant in the denominator.




It is straightforward to read the longitudinal and Hall
conductivities from Eq. (20), given by \bqa && \sigma_{L}^{c}(T) =
\mathcal{C}_{c} e^{2} v_{F}^{c 2} \rho_{c} \tau_{sc}(T) \frac{
[\tau_{sc}^{b}(T)]^{-1} + [\tau_{tr}^{a}(T)]^{-1} }{
[\tau_{tr}^{b}(T)]^{-1} + [\tau_{tr}^{a}(T)]^{-1} } , \nn &&
\sigma_{H}^{c}(T) = \mathcal{C}_{c} e^{2} v_{F}^{c 2} \rho_{c}
[\tau_{sc}(T)]^{2} \Bigl( \frac{ [\tau_{sc}^{b}(T)]^{-1} +
[\tau_{tr}^{a}(T)]^{-1} }{ [\tau_{tr}^{b}(T)]^{-1} +
[\tau_{tr}^{a}(T)]^{-1} } \Bigr)^{2} \omega_{c} . \eqa
\end{widetext}
It is clear that an additional factor $\frac{
[\tau_{sc}^{b}(T)]^{-1} + [\tau_{tr}^{a}(T)]^{-1} }{
[\tau_{tr}^{b}(T)]^{-1} + [\tau_{tr}^{a}(T)]^{-1} }$ originates
from vertex corrections. If vertex corrections are neglected in
Eq. (18), we obtain \cite{QBE_KB} \bqa && \Lambda_{c}(k_{F}
,\omega) \approx e E \frac{\Gamma_{c}^{b}(k_{F} ,\omega)}{ 2
\Gamma_{c}^{b}(k_{F} ,\omega) + i \omega_{c} } A_{c}(k_{F}
,\omega) . \nonumber \eqa Then, we reach \bqa &&
\sigma_{L}^{KB}(T) = \mathcal{C}_{c} e^{2} v_{F}^{c 2} \rho_{c}
\tau_{sc}(T) , \nn && \sigma_{H}^{KB}(T) = \mathcal{C}_{c} e^{2}
v_{F}^{c 2} \rho_{c} [\tau_{sc}(T)]^{2} \omega_{c} , \nonumber
\eqa recovering conventional expressions for both longitudinal and
transverse transport coefficients.

Resorting to the Ioffe-Larkin composite rule
\cite{Lee_Nagaosa,IL_Rule}, one can show that electrical transport
coefficients are given by contributions from conduction electrons
only \cite{Kim_TR}. Each time scale for relaxation and transport
in Eq. (21) has been evaluated in the $z = 3$ regime as follows
\cite{KB_Indranil,KB_Pepin} \bqa && [\tau_{sc}(T)]^{-1} \propto T
\ln T , ~~~ [\tau_{sc}^{b}(T)]^{-1} \propto T \ln T , \nn &&
[\tau_{tr}^{b}(T)]^{-1} \propto T^{5/3} , ~~~
[\tau_{tr}^{a}(T)]^{-1} \propto T^{5/3} . \eqa An essential aspect
in the vertex part of Eq. (21) is the presence of two competing
time scales, given by $[\tau_{sc}^{b}(T)]^{-1}$ and
$[\tau_{tr}^{a}(T)]^{-1}$, respectively. Since the temperature
dependence of $[\tau_{sc}^{b}(T)]^{-1}$ differs from that of
$[\tau_{tr}^{a}(T)]^{-1}$, we find a crossover temperature
$T_{cr}$, identified with $[\tau_{sc}^{b}(T_{cr})]^{-1} =
[\tau_{tr}^{a}(T_{cr})]^{-1}$, below which
$[\tau_{sc}^{b}(T<T_{cr})]^{-1} > [\tau_{tr}^{a}(T<T_{cr})]^{-1}$
is satisfied to result in \bqa && \rho_{QC}(T) \propto T^{5/3} .
\eqa In other words, vertex corrections are relevant to turn the
$T$-linear resistivity into the $T^{5/3}$ behavior at the Kondo
breakdown quantum critical point. On the other hand,
$[\tau_{sc}^{b}(T>T_{cr})]^{-1} < [\tau_{tr}^{a}(T>T_{cr})]^{-1}$
results above the crossover temperature $T_{cr}$, allowing the
$T$-linear behavior in the electrical resistivity because
$T^{5/3}$ contributions are cancelled in the vertex part.

An important question is the actual value of $T_{cr}$. Resorting
to the full expressions for $[\tau_{sc}^{b}(T)]^{-1}$ and
$[\tau_{tr}^{a}(T)]^{-1}$ in Ref. \cite{Kim_TR}, one can find \bqa
&& T_{cr} \approx \frac{8}{\pi k_{F}^{c2}}
\frac{\sqrt{k_{F}^{c}/k_{F}^{f}}}{m_{f}} , \eqa where
$k_{F}^{c(f)}$ is the Fermi momentum of conduction electrons
(spinons) and $m_{f}$ is the band mass of spinons. The presence of
$m_{f}$ in the denominator implies that the $T$-linear resistivity
will be observed in a wide range of temperatures because $T_{cr}$
is rather low due to $m_{f}$.

In order to make our scenario more complete, it is necessary to
compare $T_{cr}$ with $E^{*}$, below which scattering with
hybridization fluctuations becomes irrelevant, causing the typical
Fermi-liquid behavior of electrical resistivity. $E^{*}$ is
determined by the band mass of spinons and the ratio of each Fermi
momentum, given by $E^{*} = \frac{0.3}{m_{f}} \Bigl( \frac{1 -
k_{F}^{c}/k_{F}^{f}}{k_{F}^{c}/k_{F}^{f}} \Bigr)^{3}$
\cite{KB_Indranil,KB_Pepin}. As a result, we obtain $T_{cr} >
E^{*}$ in the case of $k_{F}^{c}/k_{F}^{f} > l_{c}$ while $T_{cr}
< E^{*}$ in the case of $k_{F}^{c}/k_{F}^{f} < l_{c}$, where $0 <
l_{c} = (k_{F}^{c}/k_{F}^{f})_{c} < 1$ is a constant determined by
the Fermi momentum of conduction electrons. Fig. 1 shows how
vertex corrections affect the temperature dependence of electrical
resistivity. Our point is that although the change of the
temperature dependence in the electrical resistivity is expected
due to vertex corrections, the most temperature region will show
the $T$-linear behavior at the Kondo breakdown quantum critical
point.

\begin{figure}
\includegraphics[width=0.45\textwidth]{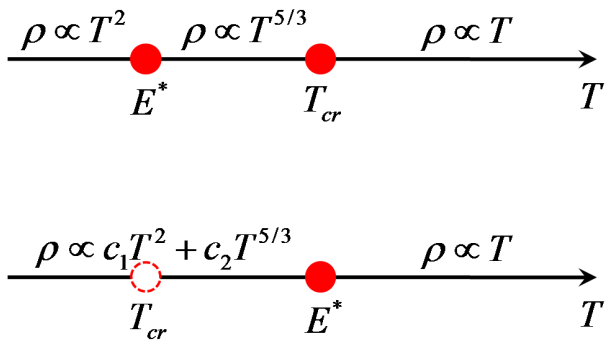}
\caption{Top : When $q^{*} = k_{F}^{f} - k_{F}^{c}$ is smaller
than a certain value, we obtain $T_{cr} > E^{*}$. Then, the
$T$-linear resistivity turns into the $T^{5/3}$ behavior from the
crossover temperature $T_{cr}$. However, the temperature regime
for the $T^{5/3}$ behavior is narrow, and the most temperature
region will show the $T$-linear resistivity. Down : When $q^{*} =
k_{F}^{f} - k_{F}^{c}$ is larger than a certain value, we obtain
$T_{cr} < E^{*}$. Then, the $T$-linear resistivity changes into
$\rho \propto c_{1} T^{2} + c_{2} T^{5/3}$, where $c_{1}$ and
$c_{2}$ are positive numerical constants, difficult to determine
with confidence in this study.} \label{fig1}
\end{figure}

It is interesting to notice that the Hall coefficient is not
renormalized due to interactions. It remains as the Fermi-liquid
value, resulting from conduction electrons only. We argue that the
non-renormalization of the Hall coefficient justifies our
calculation based on the quantum Boltzmann equation approach.
%
%


\section{Summary}

In this paper we claim that vertex corrections associated with
hybridization fluctuations are relevant in electrical transport.
Such corrections turn out to change the $T$-linear resistivity
into $T^{5/3}$ at the Kondo breakdown quantum critical point.
However, we find that the crossover temperature is too low to
clarify the regime of the $T^{5/3}$ behavior experimentally. In
other words, the $T$-linear resistivity will be observed in a wide
range of temperatures at the Kondo breakdown quantum critical
point. We justify our approximation scheme in the quantum
Boltzmann equation approach, based on the fact that the Hall
coefficient is not renormalized by interactions because the Hall
conductivity is renormalized in the same way as the longitudinal
conductivity.


We would like to close our paper, discussing the structure of our
quantum Boltzmann equation approach. An essential approximation is
that two bosonic degrees of freedom, hybridization and gauge
fluctuations, are in equilibrium. In principle, this approximation
is not consistent with fermion dynamics out of equilibrium because
such bosonic excitations originate from complex textures of
interacting fermions \cite{Maslov}. A more complete treatment will
be to start from two coupled quantum Boltzmann equations derived
from the fermion-only model after integrating over both
hybridization and gauge fluctuations. However, the time scale for
bosons to relax into equilibrium may be much shorter than that of
fermions, justifying the present approximation. Of course, this is
not an easy problem, pursued in the future. In addition, our
quantum Boltzmann equation approach seems to take ladder-type
vertex corrections in the two-band model. Recall that the quantum
Boltzmann equation approach in the one-band model introduces the
ladder type of vertex corrections, making the Ward identity
satisfied when the fermion self-energy is evaluated in the
Eliashberg approximation \cite{YB_QBE}. It will be an interesting
problem to identify relevant diagrams for our vertex corrections
in electrical transport coefficients.

This work was supported by the National Research Foundation of
Korea (NRF) grant funded by the Korea government (MEST) (No.
2011-0074542).

\end{document}